\documentclass[conference,a4paper]{APSIPA2021}
\usepackage{multirow}
\usepackage[dvips]{graphicx}
\usepackage{amsmath}
\usepackage[psamsfonts]{amssymb}
\usepackage{amsxtra}
\usepackage{threeparttable}
\usepackage{float}
\usepackage{amsfonts}
\usepackage{graphicx}
\usepackage{epstopdf}
\usepackage{array}
\newcolumntype{P}[1]{>{\centering\arraybackslash}p{#1}}

\begin{document}

\title{Incorporating Multi-Target in Multi-Stage Speech Enhancement Model for Better Generalization}

\author{%
\authorblockN{%
Lu Zhang\authorrefmark{1}, Mingjiang Wang\authorrefmark{1}, Andong Li\authorrefmark{2}, Zehua Zhang\authorrefmark{1} and Xuyi Zhuang\authorrefmark{1}
}
\authorblockA{%
\authorrefmark{1}
Harbin Institute of Technology, Shenzhen, China\\
E-mail: \{18B952047,19S052011,19S052014\}@stu.hit.edu.cn, mjwang@hit.edu.cn}
\authorblockA{%
\authorrefmark{2}
Institute of Acoustics, Chinese Academy of Sciences, Beijing, China\\
E-mail: liandong@mail.ioa.ac.cn}
}

\maketitle
\thispagestyle{empty}

\begin{abstract}
 Recent single-channel speech enhancement methods based on deep neural networks (DNNs) have achieved remarkable results, but there are still generalization problems in real scenes. Like other data-driven methods, DNN-based speech enhancement models produce significant performance degradation on untrained data. In this study, we make full use of the contribution of multi-target joint learning to the model generalization capability, and propose a lightweight and low-computing dilated convolutional network (DCN) model for a more robust speech denoising task. Our goal is to integrate the masking target, the mapping target, and the parameters of the traditional speech enhancement estimator into a DCN model to maximize their complementary advantages. To do this, we build a multi-stage learning framework to deal with multiple targets in stages to achieve their joint learning, namely ‘MT-in-MS’. Our experimental results show that compared with the state-of-the-art time domain and time-frequency domain models, this proposed low-cost DCN model can achieve better generalization performance in speaker, noise, and channel mismatch cases.
\end{abstract}

\section{Introduction}
Speech enhancement (SE), which aims at removing noise interference in the received speech signals, is a necessary signal processing technology in many real-world applications, like hearing aids, mobile phones, and automatic speech recognition (ASR) systems. It can work on the embedded devices or in the cloud, which deals with noise reduction in some simple or complex acoustic scenes. Despite the effective performance of SE, it is still challenging to design high-robustness, low-complexity methods, even real-time systems to meet the requirements of front-end processing.

Early single-channel SE approaches attempt to derive the gain functions to suppress the noise components based on some prior assumptions, such as the Gaussian distribution hypothesis and minimum mean square error (MMSE) criterion. For these traditional SE methods, such as Wiener filtering \cite{1}, MMSE-based log-spectral amplitude (MMSE-LSA) estimator \cite{2}, and the optimally modified log-spectral amplitude (OMLSA) estimator\cite{3}, accurate estimation of the noise power spectrum density (NPSD) and the prior signal-to-noise ratio (SNR) is heavily required. Despite that many traditional algorithms have been proposed to estimate the NPSD and prior SNR and validated to be robust when some ideal assumptions are satisfied, they often fail to work in some more complex non-stationary scenarios.

By dint of the powerful capability of deep neural networks (DNNs), DNN-based SE algorithms have made great progress in recent years, especially in dealing with non-stationary noises. Formulated as a supervised learning problem, direct mapping \cite{4,5}, mask \cite{6,7}, and raw waveforms\cite{8,9} are usually regarded as the targets for network training. Despite the impressive effect of DNN-based SE approaches, they usually exhibit worse robustness over traditional methods in untrained cases. Recently, recurrent units, like long short-term memory (LSTM) or gated recurrent unit (GRU), are adopted and reported to improve the generalization performance toward different noises and speakers \cite{10,11,12}. This is because the recurrent mechanism can harness the contextual information of speech utterance, which is proved to be significant for speech analysis. Besides, full convolutional models with superimposed dilated convolutions are also recommended due to their efficient long-term temporal modeling ability \cite{13,14,15}. More recently, to take advantage of the robustness of traditional methods, DeepMMSE is proposed \cite{16}, where a temporal convolutional network (TCN) is utilized to estimate the prior SNR and then integrates it into the traditional method. However, the performance is still limited in low SNRs, and the robustness in cross-corpus (channel mismatch) cases has not been explored yet.

\begin{figure*}[t]
    \centering
    \includegraphics{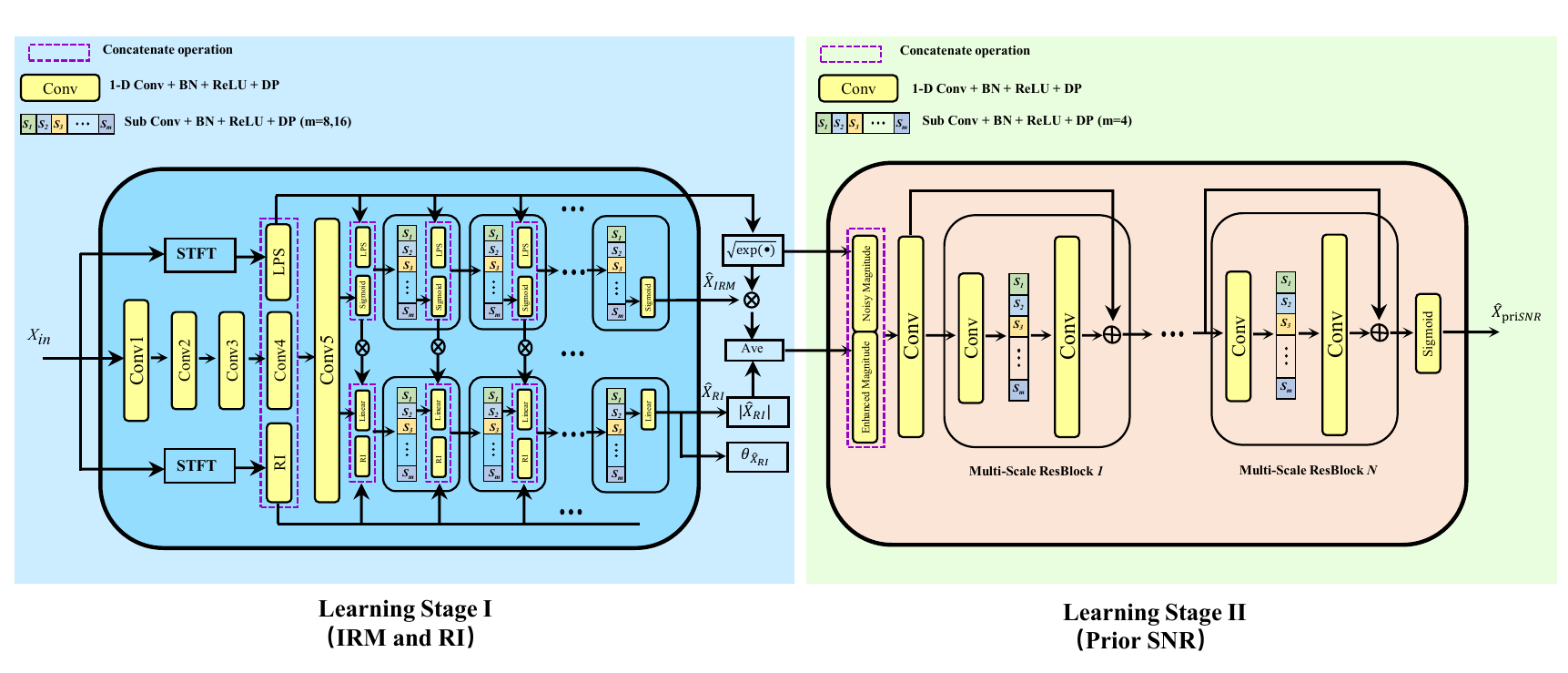}
    \caption{Speech enhancement framework of the proposed MT-in-MS model.}
    \label{fig:model}
    %\vspace{-0.5cm}
\end{figure*}

In this paper, we studied the contributions of different targets toward the generalization ability of DNNs and proposed a two-stage paradigm with multi-targets to achieve more comprehensive robustness in noise, speaker, and channel mismatch cases. Specifically, in the first stage, we proposed a novel multi-target architecture, where both complex spectrum and ideal ratio mask (IRM) are jointly learned. Based on that, the enhanced magnitude is extracted and fed into the second stage to estimate the prior SNR. The rationale is three-fold. Firstly, the phase and magnitude can be simultaneously enhanced in the first stage. Secondly, the first stage provides an SNR-improved magnitude as prior information and therefore facilitates the more accurate prior SNR estimation in the second stage, which guarantees better robustness in adverse conditions. Moreover, we fuse different targets as the final target to achieve their complementary advantages, which is helpful for model generalization. To make our model more efficient for training and inference, we only used 1-D convolutions to build a fully convolutional network model. The adopted multi-scale dilated convolution effectively reduces the model size and the required floating-point of operations (FLOPs).

The remainder of this paper is structured as follows. The formulation of multi-targets is presented in Section 2. Section 3 describes the multi-stage learning framework. Section 4 and Section 5 are the experiment results and conclusion.

\section{The Formulation of Multi-Target}

Short-time Fourier transform (STFT) is still the most effective time-frequency (TF) analysis tool for speech enhancement tasks, and it is easier to distinguish speech and noise in the TF domain than the time domain. Let \emph{y(t)} , \emph{x(t)}, and \emph{n(t)} denote the noisy speech, clean speech, and background noise, respectively, and the noisy speech \emph{y(t)} follows the additive noise model. The STFT spectrum of signals can be expressed as follows:
\begin{equation}
    Y(k,t)=X(k,t)+N(k,t)
  \label{eq1}
  %\vspace{-0.2cm}
\end{equation}
where \emph{k} and \emph{t} indicate the frequency and frame index, \emph{Y(k,t)}, \emph{X(k,t)} and \emph{N(k,t)} respectively refer to the STFT components of noisy, clean, and noise. 

As a generally recognized effective masking target, IRM is defined as follows:
\begin{equation}
    I\!R\!M(k,t)=\sqrt{\frac{X^{2}(k, l)}{X^{2}(k, l)+N^{2}(k, l)}}
    \label{eq2}
\end{equation}
The form of IRM is very similar to the square-root Wiener filter, and its data range is also from [0,1], which is easy to learn and its performance is relatively stable.

Considering phase information can help to improve speech quality, we choose complex spectrum as our mapping target to model:
\begin{equation}
    \begin{aligned}
    Y(k, l)=&[\operatorname{Re}(X(k, l))+\operatorname{Re}(N(k, l))] \\
    &+i \cdot[\operatorname{Im}(X(k, l))+\operatorname{Im}(N(k, l))] \\
    =& \operatorname{Re}(Y(k, l))+i \cdot \operatorname{Im}(Y(k, l))
    \end{aligned}
    \label{eq3}
\end{equation}
where $\operatorname{Re}(\cdot)$ and $\operatorname{Im}(\cdot)$ represents the real and imaginary parts (RI) of STFT spectrum. The RI spectrum not only has a clear harmonic structure similar to the magnitude spectrum, but also implicitly contains the phase information. Thus, it helps to reduce the phase distortion problem caused by noise.

To fuse DNN-based methods and traditional methods, the prior SNR is the key bridge. The definition of prior SNR is as follows:
\begin{equation}
    \xi(k,l)=\frac{\lambda_{x}(k,l)}{\lambda_{n}(k,l)}
    \label{eq4}
\end{equation}
where $\lambda_{x}\left(k,l\right)$=$\mathrm{E}\left\{\left|X\left(k,l\right)\right|^{2}\right\}$ and $\lambda_{n}\left(k,l\right)$=$\mathrm{E}\left\{\left|N\left(k,l\right)\right|^{2}\right\}$ represent the variance of the clean speech and noise spectrum. Since the data range of the prior SNR is too large, it is not easy to converge. Therefore, a cumulative distribution function (CDF) compression method recommended by \cite{17} is used to adjust the range to [0,1].
\begin{table}[t]
        \caption{The detailed configurations of the proposed model, \uppercase\expandafter{\romannumeral1} and \uppercase\expandafter{\romannumeral2} represent the structures of the first and second stages, respectively}
        \label{tab:The detailed}
        \centering
        \begin{tabular}{|c|c|c|c|}
            \hline
            \rule[-5pt]{0pt}{16pt}Layers  &Kernel Size    &Output Size    &Dilated Rate\\
            \hline
             \rule[-4pt]{0pt}{12pt}\uppercase\expandafter{\romannumeral1}: Conv $1$&$320\!\times \!3$&$(-1,320)$&$1$\\
            \hline
             \rule[-4pt]{0pt}{12pt}\uppercase\expandafter{\romannumeral1}: Conv $2$&$320\!\times \!3$&$(-1,161)$&$3$\\
            \hline
             \rule[-4pt]{0pt}{12pt}\uppercase\expandafter{\romannumeral1}: Conv $3$&$161\!\times \!3$&$(-1,161)$&$5$\\
            \hline
             \rule[-4pt]{0pt}{12pt}\uppercase\expandafter{\romannumeral1}: Conv $4$&$161\!\times \!1$&$(-1,161)$&$1$\\
            \hline
             \rule[-4pt]{0pt}{12pt}\uppercase\expandafter{\romannumeral1}: Conv $5$&$644\!\times \!1$&$(-1,644)$&$1$\\
            \hline
             \rule[-4pt]{0pt}{12pt}\uppercase\expandafter{\romannumeral1}: Sigmoid&$322\!\times \!1$&$(-1,161)$&$1$\\
            \hline
             \rule[-4pt]{0pt}{12pt}\uppercase\expandafter{\romannumeral1}: Linear&$644\!\times \!1$&$(-1,322)$&$1$\\
            \hline
             \rule[-4pt]{0pt}{12pt}\uppercase\expandafter{\romannumeral1}: Sub Conv&$40/41/80/81\!\times \!3$&$(-1,40/41)$&$(1,3,5)$\\
            \hline
             \rule[-4pt]{0pt}{12pt}\uppercase\expandafter{\romannumeral2}: Wide Conv&$322/161\!\times \!1$&$(-1,322)$&$1$\\
            \hline
             \rule[-4pt]{0pt}{12pt}\uppercase\expandafter{\romannumeral2}: Narrow Conv&$322\!\times \!1$&$(-1,161)$&$1$\\
            \hline
             \rule[-4pt]{0pt}{12pt}\uppercase\expandafter{\romannumeral2}: Sub Conv&$40/41/80/81\!\times \!3$&$(-1,40/41)$&$(2^d, d\!=\!0\!\sim\!4)$\\
            \hline
             \rule[-4pt]{0pt}{12pt}\uppercase\expandafter{\romannumeral2}: Sigmoid&$322\!\times \!1$&$(-1,161)$&$1$\\
            \hline
        \end{tabular}
        %\vspace{-0.4cm}
\end{table}

\section{Multi-stage Learning Model}

In order to better integrate multiple targets, we propose a two-stage learning model, and the block diagram of this model is shown in Fig.\ref{fig:model}. The detailed configurations of the model structure have been presented in Table~\ref{tab:The detailed}. We will introduce the design details in the following subsections.

\subsection{Stage I: Joint learning of IRM and RI}

As defined in Section 2, IRM characterizes the degree of noise suppression at different frequency points. That is, if there are no speech components at some frequency points, the expected IRM should be zero, while at some frequency points where the proportion of speech components is high, the value of IRM tends to be one. Therefore, it is natural to regard it as a gating operation to distinguish speech and non-speech components.

Based on this, we propose an interactive learning method between the IRM and RI targets. As shown in Fig.\ref{fig:model}, we use two parallel branches to learn IRM and RI targets interactively, and in each branch, IRM and RI targets are learned step by step through many multi-scale analysis units. For each multi-scale unit, our bidirectional multi-scale convolution method \cite{18} is adopted to encode the features. The input features are firstly divided into \textit{m} groups in multi-scale layer for 1-D sub-band convolutional operation, and then batch normalization (BN) \cite{19}, ReLU, and dropout (DP) \cite{20}. After the multi-scale layer, we respectively use a ‘linear’ and a ‘sigmoid’ layer to produce the intermediate representations of RI and IRM. Both layers use the 1-D convolution to integrate the analysis results, but the convolutional layer of the IRM branch adds the Sigmoid activation function to act on different RI feature dimensions to realize the information interaction.

Besides, since the feature dimensions of the ‘sigmoid’ and ‘linear’ layers are not equal, which are 161 dimensions and 322 dimensions respectively, we duplicate the generated gating factors and multiply them on the real and imaginary features respectively. In our model, the number of multi-scale units is set to 3 with the dilation rate of 1, 3, and 5, and the divided subgroups for IRM and RI branches are 8 and 16, respectively.

In order to extract more comprehensive acoustic features, we firstly extract the long-term features from the time-domain signals through four dilated convolution layers and then fuse the extracted TF features, log power spectrum (LPS) and RI spectrum, through a 1-D convolution layer. The noisy LPS and RI are stacked forward to the multi-scale units of two branches, which effectively shortens the gradient propagation path. At the end of the first stage model, the enhanced RI and IRM are obtained. At the same time, a simple fusion is made on the output of the first stage to prepare for the input of the second stage. For the output of the IRM branch, the estimated IRM value is multiplied by the noisy magnitude spectrum to get the masked magnitude:
\begin{equation}
    \left|\hat{X}_{I\!R\!M}(k,t)\right|=\sqrt{\exp\left(\operatorname{LPS}\left(X_{in}(k,t)\right)\right)} \cdot \hat{X}_{I\!R\!M}(k,t)
    \label{eq5}
\end{equation}
Then, we average the masked magnitude spectrum with the reconstructed magnitude from the enhanced RI:
\begin{equation}
    \left|\hat{X}(k,t)\right|=\frac{1}{2}\left(\left|\hat{X}_{R\!I}(k,t)\right|+\left|\hat{X}_{I\!R\!M}(k, t)\right|\right)
    \label{eq6}
\end{equation}
To better estimate the prior SNR, we concatenate the averaged and noisy magnitude as the input of the second stage model.

\subsection{Stage II: Learning Prior SNR}

For a priori SNR modeling scheme, previous work has proved that the TCN model [16] has more advantages than the LSTM model \cite{17} in terms of performance and computational complexity. Thus, we also adopt the TCN structure to estimate the prior SNR. The difference is that we apply the multi-scale dilated convolution used in the first stage to replace the normal 1-D dilated convolution in the original TCN residual block. This introduced multi-scale convolution layer not only reduces the number of residual blocks, but also improves the granularity of temporal feature analysis. In our proposed TCN model, we stack 8 multi-scale residual blocks, and the dilation rate cycles in increments of 1, 2, 4, 8, and 16.

The training target of the second stage model is the prior SNR, as described in Section 2. Therefore, as shown in Fig.\ref{fig:model}, after the last multi-scale residual block, a ‘sigmoid’ layer is used to estimate a compressed version of prior SNR. Different from the previous prior SNR mapping schemes, which only estimate from the noisy magnitude spectrum, the combination of enhanced magnitude and noisy magnitude can provide a better reference for the model to map prior SNR. Also, in the model inference stage, we adopt the phase spectrum recovered from the enhanced RI spectrum of the first stage to resynthesize, which can further improve the upper limit of performance.

\subsection{Loss Function}

The whole model is trained in two stages, and an intermediate estimate is obtained for each stage. Therefore, we define an accumulated loss to jointly optimize the targets of two stages:
\begin{equation}
    \text {Loss}=\alpha \cdot \text {Loss}_{\text{stage\uppercase\expandafter{\romannumeral1}}}+\beta \cdot \text {Loss}_{\text {stage\uppercase\expandafter{\romannumeral2}}}
    \label{eq7}
\end{equation}
Where $\alpha$ and $\beta$  are the weighted coefficients for each stage. The loss in the first stage is composed of two parts, which are the mean square error (MSE) loss of IRM and RI:
\begin{equation}
    \begin{aligned}
        \text {Loss}_{\text{stage \uppercase\expandafter{\romannumeral1}}}=& \frac{1}{T\cdot K}\sum_{l=1}^{T} \sum_{k=1}^{K}\left\|\hat{X}_{I\!R\!M}(k,l)\!-\!X_{I\!R\!M}(k,l)\right\|_{2}^{2} \\
        &+\frac{1}{T\cdot K}\sum_{l=1}^{T}\sum_{k=1}^{K}\left\|\hat{X}_{R\!I}(k,l)\!-\!X_{R\!I}(k, l)\right\|_{2}^{2}
    \end{aligned}
    \label{eq8}
\end{equation}
Where \(T\) and \(K\) represent the mini-batch size and feature size, respectively, and \(X_{I\!R\!M}\) and \(X_{R\!I}\) are the ideal IRM and RI targets. For the loss in the second stage, cross-entropy is used as the loss function:
\begin{equation}
    \begin{aligned}
        \text {Loss}_{\text {stage \uppercase\expandafter{\romannumeral2}}}=-&\frac{1}{T \cdot K} \sum_{l=1}^{T} \sum_{k=1}^{K}
         \left[X_{priS\!N\!R} \log \left(\hat{X}_{priS\!N\!R }\right)\right. \\
        &\left.+\left(1-X_{priS\!N\!R}\right) \log \left(1-\hat{X}_{priS\!N\!R}\right)\right]
    \end{aligned}
    \label{eq9}
\end{equation}
Where \(X_{priS\!N\!R}\) is the ideal prior SNR, log$(\cdot)$ is a logarithmic operation with base 2. Because the loss values of the two stages are of the same order of magnitude, we set $\alpha$=$\beta$=$1$ in this study.

\subsection{Multi-Target Fusion for Model Inference}

In the model inference stage, we propose a multi-target fusion method to achieve the complementary advantages of the three targets. Firstly, in order to utilize the advantages of traditional speech enhancement methods, the prior SNR predicted by the second stage model is substituted into the gain function of MMSE-LSA:
\begin{equation}
    G\!_{M\!M\!S\!E\!-\!L\!S\!A\!}(\!k,t\!)=\frac{\xi(\!k,t\!)}{\xi(\!k,t\!)\!+\!1}\!\exp \!\left\{\!\frac{1}{2}\!\int_{v(k,t)}^{\infty}\!\frac{e^{-x}}{x}dx\!\right\}
    \label{eq10}
\end{equation}
and $v(k, t)$ is given by:
\begin{equation}
    v(k, t)=\frac{\xi(k, t)}{\xi(k, t)+1} \gamma(k, t)
    \label{eq11}
\end{equation}
Where the posterior SNR $\gamma(k,t)$ is estimated by 1+$\xi(k,t)$. Then, we can obtain the MMSE-LSA enhanced magnitude:
\begin{equation}
    \left|\!\hat{X}\!_{priS\!N\!R}(\!k,\!t\!)\right|\!\!=\!\!\sqrt{\!\exp\!\left(L\!P\!S\!\left(X\!_{in}(\!k,\! t\!)\right)\right)} \!\cdot\!G\!_{M\!M\!S\!E\!-\!L\!S\!A}(\!k,\!t\!)
    \label{eq12}
\end{equation}

Moreover, the enhanced results of the first stage model are also incorporated:
\begin{equation}
    \left|\hat{X}(\!k,\!t\!)\right|\!\!=\!\!\frac{1}{3}\!\left(\left|\hat{X}\!_{R\!I}(\!k,\! t\!)\right|\!\!+\!\!\left|\hat{X}\!_{I\!R\!M}(\!k,\! t\!)\right|\!\!+\!\!\left|\hat{X}_{priS\!N\!R}(\!k,\!t\!)\right|\right)
    \label{eq13}
\end{equation}
At the same time, the reconstructed phase of the enhanced RI spectrum is used to replace the noisy phase for resynthesis:
\begin{equation}
    \theta_{\hat{X}_{R\!I}}(k,t)=\arctan \left(\frac{\operatorname{Im}\left(\hat{X}_{R\!I}(k, t)\right)}{\operatorname{Re}\left(\hat{X}_{R\!I}(k, t)\right)}\right)
    \label{eq14}
\end{equation}

\begin{equation}
    \hat{X}(k,t)=|\hat{X}(k,t)| \cdot \exp \left(i \theta_{\hat{X}_{R\!I}}(k,t)\right)
    \label{eq15}
\end{equation}
It is helpful to reduce the phase distortion caused by noises. The final enhanced waveform can be obtained through inverse STFT operations.

%\newpage
\section{Experiment and Results}

\subsection{Experimental setups}
In our experiment, we perform the model evaluation on the TIMIT speech database \cite{21}, which contains a total of 6300 sentences spoken by 630 speakers. 4,620 training utterances are corrupted by 12 noise cases (\textit{babble, factory1, destroyer1, destroyer2, cockpit1, cockpit2, volvo, tank, leopard, white, hfchannel, machinegun}) from NOISEX-92 database \cite{22} to generate a 38-hour training dataset. Each utterance is mixed with the first 60\% part of each noise file, and the mixed SNR level follows the uniform distribution in the range of -5 to 15. To guide model training, we also generate a validation dataset by mixing 280 utterances from the TIMIT test set with the middle 20\% of each noise file. As for the construction of the noisy test dataset, we consider speaker, noise, and channel mismatch to evaluate the generalization of the model. For the speaker mismatch case, 320 utterances from different speakers are mixed with the last 20\% of the 12 noises as the speaker mismatch set. Besides, 3 new noises (\textit{cockpit3, factory2, pink}) from NOISEX-92 are mixed with these 320 utterances as the noise mismatch set. To construct the more challenging channel mismatch case, we randomly select 320 utterances from Voice Bank database \cite{23} and mix them with four real noises (\textit{pedestrian area, street, bus, and café}) of CHiME-4 \cite{24}. The test SNR levels are fixed at \{-5, 0, 5, 10, 15\} dBs. The short-time objective intelligibility (STOI) \cite{25} and perceptual evaluation of speech quality (PESQ) \cite{26} are adopted as two evaluation metrics.

In our experiment, all the utterances are resampled to 16 kHz. In order to reduce the time-delay as much as possible, we split the utterance into frames by a Hamming window with 20 ms length and 10ms overlap. 320-point STFT is performed for TF analysis, and the extracted feature sizes of the waveform, LPS, and RI spectrum are 320, 161, and 322, respectively. The model is optimized by Adam \cite{27} for each mini-batch size of 10,000 consecutive frames. The dropout rate in our model is set to 0.2.

\subsection{The contributions of targets for model generalization}

\begin{figure}[h]
    \centering
    \includegraphics{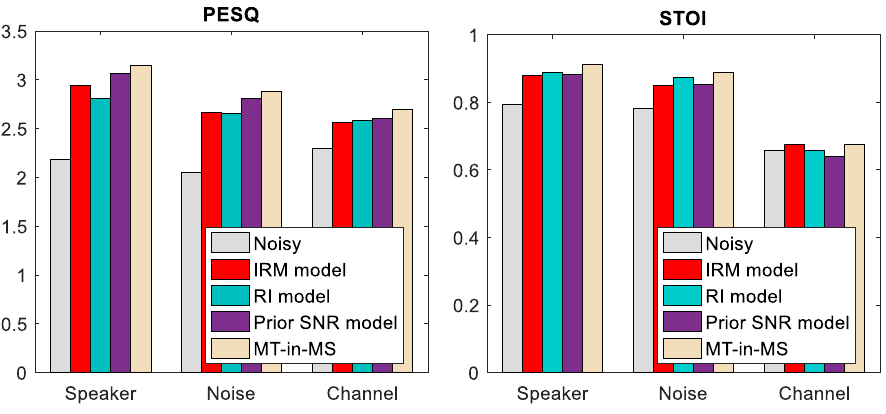}
    \caption{Averaged PESQ and STOI results for different targets in speaker, noise and channel mismatch cases.}
    \label{fig:results}
    %\vspace{-0.1cm}
\end{figure}

In this section, we study the performance of three targets (IRM, RI, and prior SNR) in the case of the speaker, noise, and channel mismatch. Three independent sub-models are separated from our MT-in-MS model, namely IRM model, RI model, and prior SNR model. Fig.\ref{fig:results} presents the averaged PESQ and STOI results under three mismatch cases. 

From the above Fig.\ref{fig:results}, we can find that the prior SNR target performs better in terms of speech quality, while the RI target produces better speech intelligibility in the speaker and noise mismatch cases. The IRM target is relatively stable in three mismatch cases and has more performance advantages in the channel mismatch case. The proposed MT-in-MS model deeply integrates the three targets, gives full play to different advantages, and achieves the best performance.

\subsection{Comparison with the state-of-the-art methods}

In this section, the comparison models are evaluated from three aspects: generalization performance, model size, and computational complexity (the floating-point of operations, FLOPs). We compare the MT-in-MS with two state-of-the-art models, GCRN \cite{11} and DDAEC \cite{9}. Both models also consider phase distortion, and process speech signals from the complex domain and time domain respectively. The evaluated results of PESQ and STOI for different models are shown in Table~\ref{tab:The averaged PESQ results for different models in speaker, noise and channel mismatch cases} and Table~\ref{tab:The averaged STOI results for different models in speaker, noise and channel mismatch cases}, respectively.
\begin{table}[th]
        \caption{The averaged PESQ results for different models in speaker, noise and channel mismatch cases}
        \label{tab:The averaged PESQ results for different models in speaker, noise and channel mismatch cases}
        \centering
        \begin{tabular}{|c|c|c|c|c|}
            \hline
            \rule[-5pt]{0pt}{16pt}{\multirow{2}{*}{Methods}} & \multicolumn{4}{c|}{\textbf{PESQ}} \\
            \cline{2-5}
             \rule[-4pt]{0pt}{12pt}& Speaker& Noise  & Channel  & Ave \\
            \hline
            \rule[-4pt]{0pt}{12pt}\rule[-4pt]{0pt}{12pt}Noisy & 2.18 & 2.05 &2.30 & 2.18\\
            \hline
            \rule[-4pt]{0pt}{12pt}GCRN \cite{11}  & 3.06  &2.85 &2.47 &2.79\\
            \hline
            \rule[-4pt]{0pt}{12pt}DDAEC \cite{9} & 2.96 &2.84 &2.58  &2.79\\
            \hline
            \rule[-4pt]{0pt}{12pt} MT-in-MS   & \textbf{3.15}  &\textbf{2.88}  &\textbf{2.70} &\textbf{2.91}\\
            \hline
        \end{tabular}
        %\vspace{-0.4cm}
\end{table}

\begin{table}[th]
    \caption{The averaged STOI results for different models in speaker, noise and channel mismatch cases}
    \label{tab:The averaged STOI results for different models in speaker, noise and channel mismatch cases}
        \centering
        \begin{tabular}{|c|c|c|c|c|}
            \hline
            \rule[-5pt]{0pt}{16pt}{\multirow{2}{*}{Methods}} & \multicolumn{4}{c|}{\textbf{STOI}} \\
            \cline{2-5}
            &\rule[-4pt]{0pt}{12pt}Speaker    &Noise  &Channel    &Ave           \\
            \hline
            \rule[-4pt]{0pt}{12pt}Noisy        &0.794         &0.781          &0.658          &0.744\\
            \hline
            \rule[-4pt]{0pt}{12pt}GCRN \cite{11}    &\textbf{0.912}&0.892          &0.645          &0.816\\
            \hline
            \rule[-4pt]{0pt}{12pt}DDAEC \cite{9}    &0.908         &\textbf{0.897} &\textbf{0.678} &\textbf{0.828}\\
            \hline
            \rule[-4pt]{0pt}{12pt}MT-in-MS     &\textbf{0.912}&0.889          &0.676          &0.826\\
            \hline
        \end{tabular}
        %\vspace{-0.4cm}
\end{table}

It can be observed that the proposed MT-in-MS model has obvious superiority in terms of PESQ, especially in the more challenging channel mismatch case. Compared with the other two methods, the proposed model has an average performance advantage of 0.12 in PESQ. As for speech intelligibility, the time-domain model DDAEC obtains better STOI results in the noise and channel mismatch cases. The GCRN model fails to improve speech intelligibility in the case of channel mismatch. Our MT-in-MS obtains STOI results very close to DDAEC. If we further consider the model size and computation, the MT-in-MS will be the optimal solution. Table~\ref{tab:Model size and FLOPs of comparison models, counted in millions (M)} has summarized the statistic model size and FLOPs of comparison models. The DDAEC and MT-in-MS models have smaller model sizes, while our model is more efficient and saves a lot of computation, thanks to the use of only 1-D convolutions.
\begin{table}[th]
    \caption{Model size and FLOPs of comparison models, counted in millions (M)}
    \label{tab:Model size and FLOPs of comparison models, counted in millions (M)}
        \centering
        \begin{tabular}{|c|c|c|c|}
            \hline
            \rule[-5pt]{0pt}{16pt}Methods &GCRN &DDAEC &MT-in-MS \\
            \hline
            \rule[-4pt]{0pt}{12pt}Model size  &9.8 M  &4.8 M    &4.8 M\\
            \hline
            \rule[-4pt]{0pt}{12pt}FLOPs       &47.7 M &776.9 M  &9.6 M\\
            \hline
        \end{tabular}
        %\vspace{-0.4cm}
\end{table}

\section{Conclusions}

Poor generalization performance is always a problem of DNN-based speech enhancement models. In this paper, we explore the generalization performance of different learning targets in various data mismatch cases, and propose a multi-target fusion scheme to achieve their complementary advantages. We use a multi-branch learning structure to integrate the IRM and RI targets, and further achieve the fusion with the traditional method in a multi-stage learning way. The incorporation of multi-scale convolution allows us to design a lightweight and efficient model to integrate different targets. Experimental results show that, compared with the state-of-the-art models, the proposed MT-in-MS model achieves better generalization performance while further reducing the computational burden for real-time processing.

\section*{Acknowledgment}

This work was supported by the Natural Science Foundation of Guangdong Province under Grant No. 2020B1515120004, and the Shenzhen Basic Research Program under Grant No. JCY20180503182125190 and JCYJ20180507182241622.

\bibliographystyle{IEEEtran}

\bibliography{mybib}

% Generated by IEEEtran.bst, version: 1.13 (2008/09/30)
\begin{thebibliography}{10}
\providecommand{\url}[1]{#1}
\csname url@samestyle\endcsname
\providecommand{\newblock}{\relax}
\providecommand{\bibinfo}[2]{#2}
\providecommand{\BIBentrySTDinterwordspacing}{\spaceskip=0pt\relax}
\providecommand{\BIBentryALTinterwordstretchfactor}{4}
\providecommand{\BIBentryALTinterwordspacing}{\spaceskip=\fontdimen2\font plus
\BIBentryALTinterwordstretchfactor\fontdimen3\font minus
  \fontdimen4\font\relax}
\providecommand{\BIBforeignlanguage}[2]{{%
\expandafter\ifx\csname l@#1\endcsname\relax
\typeout{** WARNING: IEEEtran.bst: No hyphenation pattern has been}%
\typeout{** loaded for the language `#1'. Using the pattern for}%
\typeout{** the default language instead.}%
\else
\language=\csname l@#1\endcsname
\fi
#2}}
\providecommand{\BIBdecl}{\relax}
\BIBdecl

\bibitem{1}
P.~Scalart \emph{et~al.}, ``Speech enhancement based on a priori signal to
  noise estimation,'' in \emph{ICASSP}, 1996, pp. 629--632.

\bibitem{2}
Y.~Ephraim and D.~Malah, ``Speech enhancement using a minimum mean-square error
  log-spectral amplitude estimator,'' \emph{IEEE Transactions on Acoustics,
  Speech, and Signal Processing}, vol.~33, no.~2, pp. 443--445, 1985.

\bibitem{3}
I.~Cohen and B.~Berdugo, ``Speech enhancement for non-stationary noise
  environments,'' \emph{Signal Processing}, vol.~81, no.~11, pp. 2403--2418,
  2001.

\bibitem{4}
Y.~Xu, J.~Du, L.-R. Dai, and C.-H. Lee, ``A regression approach to speech
  enhancement based on deep neural networks,'' \emph{IEEE/ACM Transactions on
  Audio, Speech, and Language Processing}, vol.~23, no.~1, pp. 7--19, 2014.

\bibitem{5}
A.~Kumar and D.~Florencio, ``Speech enhancement in multiple-noise conditions
  using deep neural networks,'' \emph{arXiv preprint arXiv:1605.02427}, 2016.

\bibitem{6}
Y.~Wang, A.~Narayanan, and D.~Wang, ``On training targets for supervised speech
  separation,'' \emph{IEEE/ACM Transactions on Audio, Speech, and Language
  Processing}, vol.~22, no.~12, pp. 1849--1858, 2014.

\bibitem{7}
D.~S. Williamson, Y.~Wang, and D.~Wang, ``Complex ratio masking for monaural
  speech separation,'' \emph{IEEE/ACM Transactions on Audio, Speech, and
  Language Processing}, vol.~24, no.~3, pp. 483--492, 2015.

\bibitem{8}
A.~Pandey and D.~Wang, ``A new framework for supervised speech enhancement in
  the time domain,'' in \emph{INTERSPEECH}, 2018, pp. 1136--1140.

\bibitem{9}
A.~{Pandey} and D.~{Wang}, ``Densely connected neural network with dilated
  convolutions for real-time speech enhancement in the time domain,'' in
  \emph{ICASSP}, 2020, pp. 6629--6633.

\bibitem{10}
J.~Chen and D.~Wang, ``Long short-term memory for speaker generalization in
  supervised speech separation,'' \emph{The Journal of the Acoustical Society
  of America}, vol. 141, no.~6, pp. 4705--4714, 2017.

\bibitem{11}
K.~Tan and D.~Wang, ``Learning complex spectral mapping with gated
  convolutional recurrent networks for monaural speech enhancement,''
  \emph{IEEE/ACM Transactions on Audio, Speech, and Language Processing},
  vol.~28, pp. 380--390, 2019.

\bibitem{12}
Y.~Hu, Y.~Liu, S.~Lv, M.~Xing, S.~Zhang, Y.~Fu, J.~Wu, B.~Zhang, and L.~Xie,
  ``{DCCRN}: Deep complex convolution recurrent network for phase-aware speech
  enhancement,'' in \emph{INTERSPEECH}, 2019, pp. 2472--2476.

\bibitem{13}
K.~Tan, J.~Chen, and D.~Wang, ``Gated residual networks with dilated
  convolutions for monaural speech enhancement,'' \emph{IEEE/ACM Transactions
  on Audio, Speech, and Language Processing}, vol.~27, no.~1, pp. 189--198,
  2018.

\bibitem{14}
S.~Pirhosseinloo and J.~S. Brumberg, ``Monaural speech enhancement with dilated
  convolutions,'' in \emph{INTERSPEECH}, 2019, pp. 3143--3147.

\bibitem{15}
A.~Li, C.~Zheng, R.~Peng, and X.~Li, ``Two heads are better than one: A
  two-stage approach for monaural noise reduction in the complex domain,''
  \emph{arXiv preprint arXiv:2011.01561}, 2020.

\bibitem{16}
Q.~Zhang, A.~Nicolson, M.~Wang, K.~K. Paliwal, and C.~Wang, ``{DeepMMSE}: A
  deep learning approach to {MMSE-based} noise power spectral density
  estimation,'' \emph{IEEE/ACM Transactions on Audio, Speech, and Language
  Processing}, vol.~28, pp. 1404--1415, 2020.

\bibitem{17}
A.~Nicolson and K.~K. Paliwal, ``Deep learning for minimum mean-square error
  approaches to speech enhancement,'' \emph{Speech Communication}, vol. 111,
  pp. 44--55, 2019.

\bibitem{18}
L.~Zhang and M.~Wang, ``{Multi-Scale TCN}: Exploring better temporal {DNN}
  model for causal speech enhancement,'' in \emph{INTERSPEECH}, 2020, pp.
  2672--2676.

\bibitem{19}
S.~Ioffe and C.~Szegedy, ``Batch normalization: Accelerating deep network
  training by reducing internal covariate shift,'' in \emph{ICML}, 2015, pp.
  448--456.

\bibitem{20}
N.~Srivastava, G.~Hinton, A.~Krizhevsky, I.~Sutskever, and R.~Salakhutdinov,
  ``Dropout: a simple way to prevent neural networks from overfitting,''
  \emph{Journal of Machine Learning Research}, vol.~15, pp. 1929--1958, 2014.

\bibitem{21}
J.~S. Garofolo, L.~F. Lamel, W.~M. Fisher, J.~G. Fiscus, and D.~S. Pallett,
  ``Getting started with the {DARPA TIMIT CD-ROM}: An acoustic phonetic
  continuous speech database,'' \emph{National Institute of Standards and
  Technology (NIST), Gaithersburgh, MD}, 1988.

\bibitem{22}
A.~Varga and H.~J. Steeneken, ``Assessment for automatic speech recognition:
  {II. NOISEX-92}: A database and an experiment to study the effect of additive
  noise on speech recognition systems,'' \emph{Speech Communication}, vol.~12,
  pp. 247--251, 1993.

\bibitem{23}
C.~Valentini-Botinhao, X.~Wang, S.~Takaki, and J.~Yamagishi, ``Investigating
  {RNN}-based speech enhancement methods for noise-robust text-to-speech.'' in
  \emph{SSW}, 2016, pp. 146--152.

\bibitem{24}
E.~Vincent, S.~Watanabe, A.~A. Nugraha, J.~Barker, and R.~Marxer, ``An analysis
  of environment, microphone and data simulation mismatches in robust speech
  recognition,'' \emph{Computer Speech \& Language}, vol.~46, pp. 535--557,
  2017.

\bibitem{25}
C.~H. Taal, R.~C. Hendriks, R.~Heusdens, and J.~Jensen, ``An algorithm for
  intelligibility prediction of time--frequency weighted noisy speech,''
  \emph{IEEE/ACM Transactions on Audio, Speech, and Language Processing},
  vol.~19, pp. 2125--2136, 2011.

\bibitem{26}
A.~W. Rix, J.~G. Beerends, M.~P. Hollier, and A.~P. Hekstra, ``Perceptual
  evaluation of speech quality {(PESQ)}-a new method for speech quality
  assessment of telephone networks and codecs,'' in \emph{ICASSP}, 2001, pp.
  749--752.

\bibitem{27}
D.~P. Kingma and J.~Ba, ``Adam: A method for stochastic optimization,'' in
  \emph{ICLR}, 2014, pp. 1--13.

\end{thebibliography}

\end{document}